\documentclass[11pt]{article}
\usepackage[utf8]{inputenc}

\usepackage{physics}
\usepackage{graphicx}
\usepackage{natbib}
\usepackage{amsmath}
\usepackage{amssymb}
\usepackage[margin=1.2in]{geometry}
\usepackage[multiple]{footmisc}
\usepackage{url}

\title{As Below, So Before: `Synchronic' and `Diachronic' Conceptions of Spacetime Emergence}
\author{Karen Crowther\footnote{Department of Philosophy, Classics, History of Art and Ideas, University of Oslo} \footnote{karen.crowther@ifikk.uio.no}}

\begin{document}
	
	\maketitle
	\bibliographystyle{newapa}
	
	\begin{abstract}
		Typically, a less fundamental theory, or structure, emerging from a more fundamental one is an example of \textit{synchronic} emergence. A model (and the physical state it describes) emerging from a prior model (state) upon which it nevertheless depends is an example of \textit{diachronic} emergence. The case of spacetime emergent from quantum gravity and quantum cosmology challenges these two conceptions of emergence. Here, I propose two more-general conceptions of emergence, analogous to the synchronic and diachronic ones, but which are potentially applicable to the case of emergent spacetime: an inter-level, hierarchical conception, and an intra-level, `flat' conception. I then explore whether, and how, these ideas may be applicable in the case of several putative examples of relativistic spacetime emergent from the non-spatiotemporal structures described by different approaches to quantum gravity, and of spacetime emergent from a non-spatiotemporal `big bang' state according to different examples of quantum cosmology.  
		
	\end{abstract}
	
	\tableofcontents
	
	\section{Introduction}
	
	Our best understanding of spacetime is provided by the theory of \textit{general relativity} (GR), which describes gravity as the curvature of spacetime due to the presence of energy and massive bodies. GR is extremely well-tested in all accessible domains, and is regarded as one of the most successful scientific theories ever. Yet, physicists do not believe that GR---along with the description of spacetime it provides---is fundamental. Instead, GR is expected to be incorrect at extremely high energy scales (short length scales), and in regions of extremely high curvature, where quantum effects cannot be neglected. The correct description of the physics in these domains is meant to be provided by a theory of \textit{quantum gravity} (QG), which is currently being sought. While we do not have an accepted theory of QG at present, there are a number of approaches towards finding such a theory (i.e., different research programs). Several of these describe physics that is radically distinct from that of GR, and which is arguably \textit{non-spatiotemporal}---lacking some important features of spacetime, space and/or time \citep{Huggett2013}. 
	
	QG is meant to describe the more fundamental physics that `underlies' GR. One view of QG is that it captures the non-spatiotemporal physics that is responsible for the appearance of gravitational phenomena in accessible domains (those where GR is well-tested). This physics is supposed to be \textit{quantum} in some sense (although it may necessitate the reformulation of quantum field theory), and so one picture of QG is that describe discrete `quanta' or `atoms' \textit{of} spacetime, that themselves do not exist \textit{in} spacetime. On this view, QG may be seen as more fundamental than GR in two senses: first, QG would be more fundamental than GR in an analogous way to how the atomic description of matter is seen as more fundamental than the continuous description of matter, and, second, QG would be more fundamental than GR in the same way that the quantum description of matter is more fundamental than the classical one. If QG is indeed radically distinct from GR in this way, then, given the remarkable success and stability of GR at all accessible energy scales\footnote{Ignoring the problem of dark matter, which may indicate a problem with GR at large length scales.} (e.g., treatment of GR as an effective field theory shows that quantum corrections are negligible in these domains\footnote{See, e.g., \citet{Burgess2004}.}), we would expect GR to \textit{emerge} from QG in the appropriate domain, i.e., at low energy scales \citep{Crowther2018}.
	
	A less-fundamental theory emerging from a more-fundamental one is an example of what would usually be called \textit{synchronic} emergence: the two theories are supposed to apply to the same systems at the same time, or otherwise under the same conditions, but at different `levels' (e.g., different energy scales). The term `synchronic' in this particular case is unfit, however, given that QG may not have an associated conception of time. A more general characterisation of `hierarchical', or inter-level, emergence is required; I provide this below (\S\ref{sync}), where I also explore several examples of this form of emergence in physics. I argue that the general account of hierarchical emergence that I present here is: 1., a \textit{bona fide} account of emergence that generally accords with what is typically understood as `emergence' in the philosophy of science literature; 2., is applicable in the case of spacetime emergent from QG; and 3., that it is likely exemplified in a number of approaches to QG, including \textit{analogue spacetime}, \textit{loop quantum gravity} and \textit{causal set theory}.
	
	The `synchronic', or hierarchical conception of emergence is not the only one that is interesting in the case of spacetime emergent from QG---there is also scope for a `diachronic' or `flat' conception of emergence, which would describe a spatiotemporal state emergent from a `prior' non-spatiotemporal state on the same level. While in the case of hierarchical emergence we consider the relationship between theories (or models) describing the system at different levels, in the case of flat emergence we consider the relationship between two different states of the system. Usually, these states would be connected causally, and distinguished temporally, but in the case of spacetime emergence this is not possible in general---hence, the name `diachronic emergence' is inapt, and a more general account of `flat', or intra-level emergence is required, which I develop below (\S\ref{diac}).
	
	The flat conception of spacetime emergence is possible because one of the domains where GR is expected to be incorrect, and to require replacement by QG, is at the very beginning of the universe. Using GR and current observations of the large-scale structure of the universe, cosmologists extrapolate backwards in time in order to produce a description of the past evolution of the universe. The resulting picture is the standard, or `big bang', model of cosmology, which describes the universe expanding from a state of extremely high temperature and density approximately 13 billion years ago. Before this, however, there is the big bang singularity. One interpretation of the singular behaviour of the model is that GR is incorrect in this domain, due to its neglect of quantum effects that become important at extreme density and temperature (in which case GR likely becomes incorrect at some finite time approaching the singularity). On this view, the singularity is an unphysical artefact---a signal that GR is inapplicable here---and thus, QG should provide a correct, non-singular description of the physics in this domain. Indeed, there are proposals along these lines: \textit{loop quantum cosmology} presents a model which is continuous through the region which classically would correspond to the big bang singularity \citep[see, e.g.,][]{Bojowald2011}. On the `other side of the big bang', the model arguably describes a state that is purely spatial, with no associated notion of time \citep{Brahma2017}.
	
	\citet{Brahma2017} argues that the transition from the purely spatial state to the spatiotemporal one in this model represents ``the emergence of time in loop quantum gravity'', while \citet{Huggett2018} claim that it represents the ``(a)temporal emergence of spacetime''. Yet, neither of these papers explain precisely what they mean by `emergence'. Here (\S\ref{diac}), I provide an account of emergence that is capable of accommodating cases of spacetime emergent from a `prior' non-spatiotemporal\footnote{Here, and throughout the paper, I follow \citet{Huggett2013, Huggett2018} in using ``non-spatiotemporal'' to refer to any theory or model that describes physics that is ``less than fully spatiotemporal in some significant regard''.} state. This account of emergence is an example of `diachronic' or flat emergence, inspired by the work of \cite{Guay2016, Sartenaer2018}, but which is much weaker and more general than their account of `transformational emergence'. The proposed account of flat emergence is analogous to the hierarchical conception of emergence that I advocate in the case of QG.
	
	I argue that the general account of flat emergence that I present here is: 1., a \textit{bona fide} account of emergence that accords with what is typically understood as `emergence' in the general philosophy literature; 2., is applicable in the case of QG; and 3., that it is potentially exemplified in a number of approaches to QG---specifially, those which describe spacetime emergent from a `prior' non-spatiotemporal state via a \textit{geometrogenesis} phase transition (\S\ref{geo}, as well as \textit{causal set cosmology} (\S\ref{CScosmo}), and possibly also the loop quantum cosmology model (\S\ref{LQC}).
	
	However, it must be emphasised from the very beginning of this paper that these QG cosmology models are far from being fully developed or understood! At present, any interpretation is reliant upon speculation, and is highly precarious. It is not clear whether these models are physically meaningful at all.
	
	I begin (\S\ref{fund}) by discussing the ways in which QG (and the physics it describes) may be considered more fundamental than GR, and the requirement that GR be reducible to QG in the relevant domain. Next (\S\ref{emerg}), I present the general account of emergence that is applicable in the case of QG, and explain how it compares to the more-familiar understanding of emergence in the philosophy literature. In (\S\ref{sync}), I discuss the ways in which spacetime may be said to emerge `hierarchically' from the physics described by QG, using the examples of analogue models of spacetime, causal set theory, and loop quantum gravity. In (\S\ref{diac}), I explore the ways in which spacetime may be said to emerge `flatly' from the physics described by QG, using the examples of causal set cosmology, pregeometric approaches to QG, and loop quantum cosmology.
	
	\section{Theory reduction and relative fundamentality in quantum gravity}\label{fund}
	
	Theory \textit{L} is said to be \textit{reducible} to theory \textit{M} if \textit{L} is deducible from \textit{M}, approximately, and in the appropriate domain (i.e., that where \textit{L} is known to be successful). Theory reduction in this sense demonstrates that \textit{M} has a broader domain than \textit{L}---i.e., that \textit{M} is capable of approximately describing all of the phenomena that \textit{L} successfully describes, plus more. It is a requirement on any acceptable theory of QG that GR be reducible to it in this sense---i.e., it is part of the definition of QG that GR be approximately derivable from QG in the domain where GR is known to be successful \citep{Crowther2018}. This means that even if QG does not describe spacetime fundamentally, the conception of spacetime described by GR should be approximately recoverable (derivable) from QG in the regimes where spacetime is known to be a successful concept.
	
	Next, consider relative fundamentality: given a particular system \textit{S}, or phenomenon \textit{P}, a \textit{more fundamental} theory, \textit{M}, is one that provides a more basic description of \textit{S} or \textit{P} than a \textit{less fundamental} theory, \textit{L}, does. There is only one condition for relative fundamentality: that the laws of \textit{L} depend (at least partly) on the physics described by \textit{M}, and not vice-versa---i.e., relative fundamentality is \textit{asymmetric dependence}.\footnote{This characterisation is supposed to be compatible with the idea of `top-down causation', depending on how the `higher level' and `lower level' labels are applied in the particular proposals (e.g., `higher level' could not refer to a less fundamental theory on this account).} This sense of relative fundamentality may be demonstrated by \textit{L} being reducible to \textit{M}, and \textit{M} not being reducible to \textit{L}---i.e., \textit{L} being approximately and appropriately derivable from \textit{M}, but not the other way around.\footnote{See \citet{Crowther2018}.}
	
	Recall that, on one view of QG, the theory is supposed to provide both a \textit{micro} description of spacetime (because the domain of necessity of QG includes extremely short length scales)\footnote{\label{foot:micro}Here, `micro' is used purely in a figurative sense, as a means of distinguishing the degrees of freedom described by QG from those (`macro' degrees of freedom) of current physics. `High-energy scales' and `short-length scales' are used interchangeably, and are also used to signify the domain expected to be described by QG. The scare quotes indicate that this may not literally be true, because QG may describe a regime where the idea of length (and, correspondingly, energy) are not meaningful.} as well as a \textit{quantum} description of spacetime. Each of these is sufficient for claiming that QG is more fundamental than GR, taking relative fundamentality as asymmetric dependence. So, QG may be seen as more fundamental than GR in two senses: first, QG would be more fundamental than GR in an analogous way to how the atomic description of matter is seen as more fundamental than the continuous description of matter, and, second, QG would be more fundamental than GR in the same way that the quantum description of matter is more fundamental than the classical one. 
	
	It is likely that the recovery of spacetime from the more fundamental structures described by QG is a two-step process, involving both a quantum/classical transition (by which we recover the `classical appearance' of spacetime), as well as a micro/macro transition (by which we arrive back at at familiar energy scales). While both these transitions address the question of why we do not need to use QG to describe much of the gravitational phenomena we observe, they are distinct, and may or may not be related to one another.\footnote{See, e.g. \cite{Butterfield1999}, as well as \cite{Oriti2018, Oriti2018a}, which provide a taxonomy as well as many concrete examples from different approaches to QG.} Both transitions represent common problems in physics and the philosophy of physics; and both play a role in understanding the relationships of emergence and reduction.
	
	The micro/macro transition is not something that happens to a system, but a change in the level of description---it is the re-framing to a coarser-grained theory. The micro/macro transition may be represented by an approximation procedure, a limiting process (such as the thermodynamic or `continuum limit'), or the \textit{renormalisation group flow} and the other methods of \textit{effective field theory} (EFT). The framework of EFT is employed in order to produce a theory valid at a certain energy scale from one valid at a different scale. For instance, this framework sets out a procedure for creating a low-energy (macro) theory from a high-energy (micro) one describing the same system (in terms of different degrees of freedom). It also provides an explanation for how it is that we can use our macro theories without needing to know the details of the micro-physics.\footnote{See, e.g., 
		\cite{Batterman2005, Batterman2011, Crowther2015, Franklin2018, Hartmann2001}.} All of these different techniques are employed by various approaches to QG, in their attempts to connect QG back to GR.
	
	The limiting and approximation procedures used to describe the micro/macro transition are incapable of resolving a quantum superposition. Since we expect the generic states described by QG to be superpositions, a treatment of the quantum/classical transition is necessary for understanding the classical appearance of spacetime.\footnote{See, e.g., the discussion in \cite{Wuthrich2017}.} Here, there are two different issues that need to be addressed. First, quantum theories are supposed to apply universally (this fact underlies one of the main motivations for QG), so there is the question of why, in practice, they are usually only necessary for describing small systems. Part of the answer to this first question may lie in the idea of \textit{decoherence}, which says that the interference effects associated with superpositions become suppressed through a system's interactions with its environment. Larger systems more strongly couple to their environments, so the interference effects rapidly become undetectable (although the system remains inherently quantum). This gives us some insight into the transition that a system undergoes that prompts us to move from a quantum description of it to a classical one. 
	
	Decoherence, however, does not give us an answer to the second issue, which is the \textit{measurement problem}: why it is that any measurement on a quantum system finds the system in a definite state even though the system evolves as a superposition of different states. It seems likely that a solution to the measurement problem is required if we are to fully understand the relationship between spacetime and the quantum structures that constitute it---or it may be that the solution will be provided by the QG itself.\footnote{Regarding this second possibility, see e.g. \cite{Penrose1999, Penrose2002}.}

	The recovery of spacetime from QG, and the reducibility of GR to QG, can be used to satisfy one of the characteristic conditions of `synchronic' emergence---that the emergent theory depend on the more basic physics. For the rest of the paper, however, I focus less on the details of this aspect, and more on the other two characteristics of emergence, being the novelty and autonomy of the emergent physics compared to its emergence basis.
	
	\section{Two conceptions of emergence}\label{emerg}
	
	Emergence is an empirical relation between two relata of the same nature, an emergent, $E$, and its basis, $B$. Depending on the case of interest, $E$ and $B$ may be objects, properties, powers, laws, theories, models, etc., as specified by the particular account of emergence being appealed to. The most general schema of emergence holds that $E$ is at once \textit{dependent} on $B$ and \textit{novel} compared to, or \textit{autonomous} from, $B$. Again, the relevant senses of `dependence', `novelty' and `autonomy' are to be specified by the particular account being used. 
	
	Here, I am interested in emergence as a relation between theories or parts of theories, such as models. I take the general conception of emergence to comprise three claims,
	
	\begin{description}
		\item[EMERGENCE: GENERAL CONCEPTION]
		\item[Dependence:] $E$ is dependent on $B$, as evidenced by $E$ being approximately derivable from $B$, and/or supervenient upon $B$ (supervenience means that there would be no change in $E$ unless there were a change in $B$, but not vice versa);
		\item[Novelty:] $E$ is strikingly, qualitatively different from $B$;
		\item[Autonomy:] $E$ is robust against changes in $B$; or $E$ is the same for various choices of, or assumptions about, the comparison class, $B$
	\end{description}
	
	This is a positive conception of emergence, which does not require novelty to be the failure of reduction, deduction, explanation, or derivation in any sense. Such a positive conception of emergence is now familiar in the philosophy of physics generally, and the philosophy of QG in particular.\footnote{Such a positive conception of emergence is now familiar in the philosophy of physics generally, and in the philosophy of QG in particular, having been employed, e.g., by \cite{Butterfield2011a, Butterfield2011b, Crowther2015, Crowther2016, deHaro2017, Dieks2015, Linnemann2018}, among others.} The positive conception of emergence is the most appropriate for understanding the case of spacetime emergent from QG for two reasons. First, as explained above, GR must be reducible to QG---i.e., it is a requirement on any theory of QG that GR be approximately and appropriately derivable from it. This condition may be used to satisfy the `Dependence' claim of emergence. Thus, we need an account of emergence that is compatible with reduction, at least in this sense. Second, none of the approaches to QG are complete, so basing any claims of emergence on their failure to explain, derive, or predict particular aspects of GR spacetime is risky, given that a central goal of each of the approaches is to develop a theory that approximately and appropriately \textit{recovers} (i.e., derives and explains) GR spacetime.

	Of course, the incomplete state of all of the approaches means that any philosophy of QG that seeks to interpret these fragmentary theories is precarious in the sense that its findings may not be relevant once the full theory is known. The positive interpretational program is, nevertheless, more interesting and worthwhile than the pessimistic one---given the aim of each approach to develop a theory that recovers GR spacetime, we can make assumptions about their success in this regard and begin to speculate on the positive ideas of emergence they may suggest, while taking into account their current, incomplete state of development. I employ this strategy when considering emergence in the examples I consider below (\S \ref{sync}--\ref{diac}).
	
	As stated, I take $E$ and $B$ to be theories, or parts of theories (e.g., models), depending on the particular case-studies (below). This may seem to commit me to a purely epistemological account of emergence, but I merely wish to remain neutral in regards to how these structures relate to the world. If the reader is interested in ontological emergence, I advocate engaging in a naturalised metaphysics, taking the ontology to be the entities putatively described by the theories (entities such as the discrete elements of causal set theory, the quanta of area and volume described by loop quantum gravity, or spacetime described by a particular model of GR, for instance). My account of emergence is intended to be a neutral one that could be interpreted as either epistemological or ontological. (But, if the reader is convinced that ontological emergence must involve a failure of reduction, then they are better to treat this as an epistemological account of emergence, and are advised not to look to spacetime emergent from QG for an example of ontological emergence, given my comments above).
	
	While it is a requirement on QG that it be more fundamental than GR, and that GR spacetime be approximately and appropriately derivable from QG, it is \textit{not} a requirement that GR spacetime be \textit{emergent} from the physics described by QG. Indeed, there are examples of approaches which likely would not meet the conditions for emergent spacetime (e.g., some approaches utilise conceptions of spacetime and plausibly do not provide a strong basis for the \textit{Novelty} claim of emergence). Yet, there are examples of approaches which can plausibly be interpreted as candidates for emergent spacetime---some of which I present in the next two sections.\footnote{While I will argue that these examples plausibly can be interpreted as candidates for spacetime emergence, however, there is also scope for arguing that we don't need to understand them as emergence, and instead adopt a different metaphysical interpretation of the relationship between spacetime and the structures described by QG; cf. \cite{LeBihan2018}. Another option is to utilise the idea of `partial functionalism', rather than metaphysical accounts of emergence, cf. \cite{Baron2019}.}
	
	As stated, there are two different potential conceptions of emergent spacetime from QG: \textit{hierarchical} emergence and \textit{flat} emergence. These are intended as more general versions of \textit{synchronic} and \textit{diachronic} conceptions of emergence, respectively. In the synchronic conception of emergence, $B$ and $E$ represent different levels of description: $B$ is said to describe the system at the \textit{lower leve}l and $E$ at the \textit{higher level}. In physics, $B$ and $E$ may be theories that apply at different ranges of length- or energy-scales, where, typically, $B$ describes the system at higher energy scales (shorter length scales) than $E$, which applies at comparatively low energy scales (large length scales). These theories are supposed to apply to the system at the same time, or otherwise under the same conditions, i.e, there is no `change' considered, except the level at which you view the system. 
	
\begin{figure}[!htb]
		\center{\includegraphics[width=\textwidth]
			{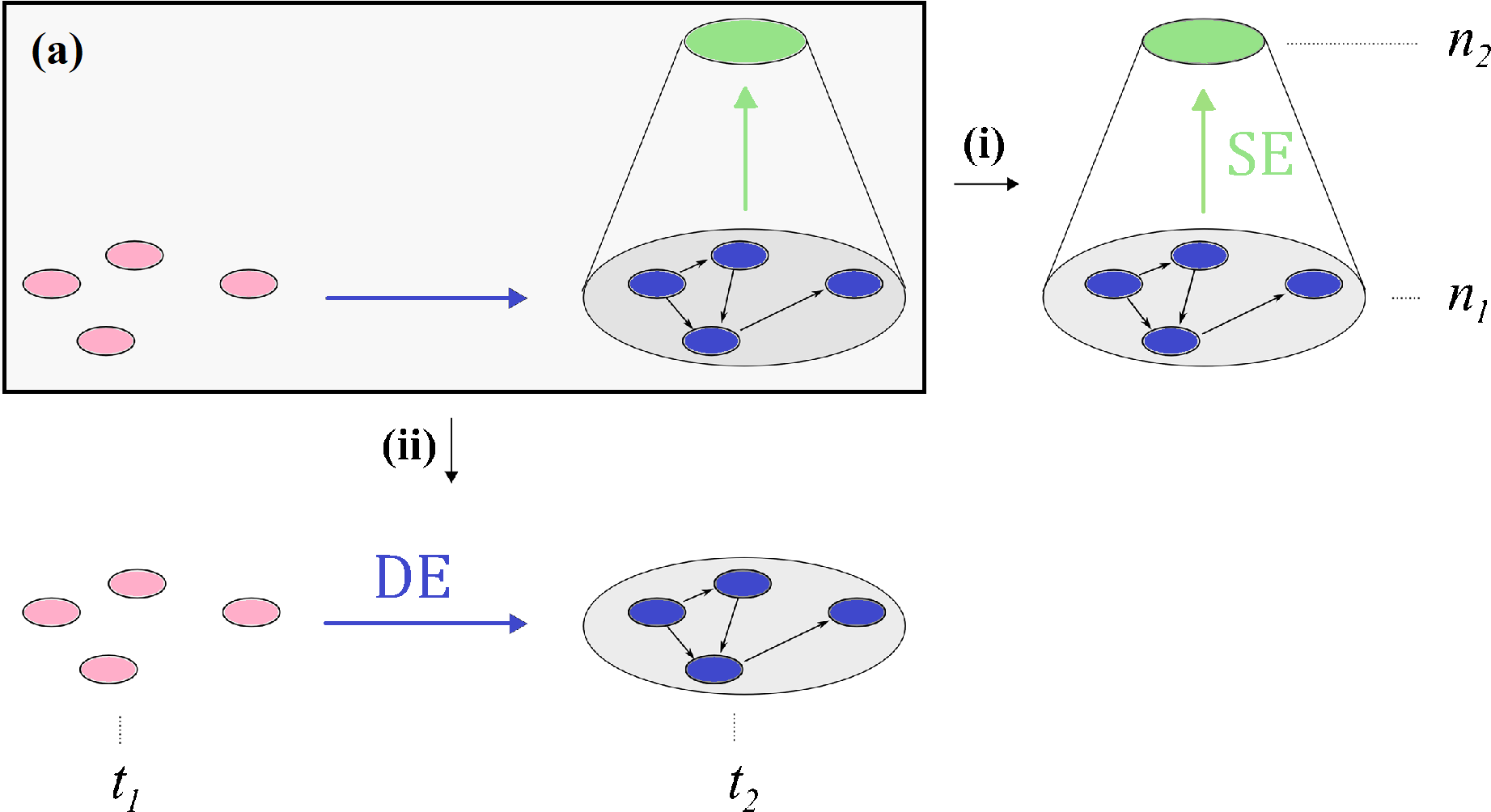}}
		\caption{\label{fig:guaysart} Two conceptions of emergence.  A system \textbf{(a)} at time $t_1$ (pink) has changed at time $t_2$ (blue), resulting in some novel higher-level phenomena (green). \textbf{(i)} If we are interested in synchronic emergence (arrow SE), we consider the system at level $n_2$ compared to the system viewed at $n_1$, at a single time, here $t_2$. \textbf{(ii)} If we are interested in diachronic emergence (arrow DE), we consider the system at $t_2$ compared to the system at $t_1$, at a single level, here $n_1$. Figure adapted from \cite{Guay2016}.}
	\end{figure}

	In the diachronic conception of emergence, $E$ and $B$ describe the system at the same level. These theories, or models, are supposed to apply to the system at different times, or otherwise under different conditions. The idea is that the system has undergone some change: typically, $B$ describes it before, and $E$ after. This conception of emergence is not associated with a notion of fundamentality. The difference between these two conceptions of emergence is illustrated in Fig.\ref{fig:guaysart}. Specific accounts and examples of synchronic and diachronic emergence are presented in the relevant sections that follow.

	\section{Hierarchical emergence of spacetime}\label{sync}
	
	The main challenge faced by the hierarchical (`synchronic') view of emergence when applied to the case of spacetime is defining the idea of `levels' without reference to length scales. While the usual way of evading this worry is by referring instead to energy scales, this does little but sweep the problem under the carpet, given that energy scales can be defined in terms of length scales. A more useful approach is to distinguish levels in terms of `size of grain'---i.e., a lower-level theory provides a finer grained description of the physics, while a higher-level theory provides a coarser-grained description of the phenomena. Typically (especially in high-energy physics), these notions correlate with the energy scales at which the theories apply, which in turn tends to correlate with the hierarchy of relative fundamentality (though these notions are not necessarily related). Thus, for convenience of notation, too, I will still distinguish the lower level physics described by $B$ as `micro' physics, and the emergent level $E$ physics as `macro' phenomena.\footnote{See fn.\ref{foot:micro}.} 
	
	A useful account of hierarchical emergence is developed by considering the framework of EFT, mentioned above---this is a `toolbox' for constructing (field) theories that are valid at low energy scales compared to a given heavy mass `cutoff', approaching which the theory starts to misbehave formally, and is no longer considered a reliable description of the system of interest. (More generally, a theory is referred to as `effective' if it is considered to be reliable only within a restricted domain---i.e., effective theories are not universally valid, and are thus usually not considered fundamental). 
	
	A simple example is analogue models of spacetime in fluids. These models can be built concretely in the physics laboratory, using familiar liquids such as water, or more exotic superfluids such as Bose-Einstein condensates (BEC). A heuristic picture of such a model, is to imagine that when the system is probed at low energy, the particles in the fluid vibrate, producing sound waves---the quanta of which are \textit{phonons} (`sound particles'), a type of \textit{quasiparticle}. These quasiparticles are subject to an effective curved-spacetime metric, meaning they behave as though they `exist in' curved spacetime, oblivious to the underlying (flat) surface of the fluid itself. As energy is increased, however, the quasiparticles eventually have short enough wavelength to `detect' the discrete particles of the fluid (i.e., when the quasiparticles have wavelengths comparable to the distances between the particles), and the effective theory used to describe the quasiparticles ceases to be valid.
	
	Very simply, we begin with a BEC with particle density $\rho$ and coherent phase $\theta$. In constructing the analogue model, these variables are linearly expanded about their ground state values, $\rho_0$ and $\theta_0$, to give $\rho = \rho_0 + \delta \rho$ and $\theta = \theta_0 + \delta \theta$, where $\delta \rho$ and $\delta \theta$ represent fluctuations in density and phase. These variables are then substituted into the Lagrangian describing the BEC, and the high-energy fluctuations are identified and `integrated out', so that only the low-energy interactions are included in the theory. The result is, schematically, a sum of two terms: $L = L_0 [\rho, \theta] + L_{\mathit{eff}}[\delta\theta]$, where $L_0$ is the Lagrangian describing the ground state of the BEC, and $L_{\mathit{eff}}$ is the effective Lagrangian describing the low energy fluctuations above the ground state. $L_{\mathit{eff}}$ is formally identical to the Lagrangian of a massless scalar field in (3+1)-dimension spacetime, and the curved effective metric depends explicitly on the velocity of the underlying fluid.\footnote{For details see, e.g., \cite{Barcelo2011} for a review, or \cite{Bain2008, Bain2013} for a description aimed at philosophers.}
	
	As \cite{Bain2013} points out, given the substantial difference between $L_0$ and $L_{\mathit{eff}}$, particularly the different symmetries they exemplify--- $L_0$ being non-relativistic (Galilei invariant), and $L_{\mathit{eff}}$ being relativistic (Lorentz invariant)---we can treat the original Lagrangian and the effective Lagrangian as describing two different theories. The analogue models show us that curved spacetime metric is incredibly easy to obtain from a variety of different systems. Formally, emergent Lorentz invariance is a generic feature of the linearisation procedure used to construct these models. Thus, given the low-energy physics, the high-energy theory is severely underdetermined. This is a hallmark of the EFT program. 
	
	Analogue relativistic spacetime can be considered an emergent phenomenon because it fits the general schema of emergence (\S\ref{emerg}), where the theory describing the quasiparticles, $L_{\mathit{eff}}$ is said to be emergent, $E$, from the basis theory, $B$, describing the ground state of the fluid, $L_0$. The \textit{Dependence} condition is satisfied because the effective theory, $E$, is constructed from the `micro' theory, $B$. The quasiparticles described by $E$ are collective phenomena: low-energy excitations of the underlying particles of the fluid (i.e., they have no independent, or fundamental, existence). \textit{Novelty} is satisfied because $E$ and $B$ are characterised by different symmetries; $E$ is Lorentz-invariant while $B$ is Galilei-invariant. Finally, \textit{Autonomy} is satisfied because effective curved spacetime is very easy to obtain from a variety of different systems, with different micro-constitutions. Given only the low-energy physics, the high-energy theory is severely underdetermined.
	
	More generally, this case-study suggests the following account of hierarchical emergence:
	
	\begin{description}
		\item[EMERGENCE: HIERARCHICAL CONCEPTION]
		\item[Dependence:] The coarser grained theory (model, or structure) $E$ is constructed (i.e., 	derived) from the finer grained theory $B$. The physics described by the laws of $E$ may be said to \textit{supervene} on those of $B$. (Supervenience means that there is no change at the $E$-level unless there is a change at the $B$-level, but not vice versa).
		\item[Novelty:] The physics described by the coarser grained, or low energy (`macro') theory $E$	differs remarkably from that of the finer grained, or higher energy (`micro') theory $B$;
		\item[Autonomy:] The physics described by $E$ is robust against changes in the micro physics; $B$ is underdetermined by $E$.
	\end{description}

	A few comments are necessary. In regards to \textit{Novelty}, note that it need not be an asymmetric relation; this condition just captures the ways in which the two theories differ from one another. In regards to \textit{Autonomy}, there are two senses of theory-underdetermination that may be relevant.\footnote{See also \cite{Franklin2018}, who distinguishes two different senses of autonomy related to EFT and theoretical naturalness.} These come about due to the universality (multiple realisability) of the $E$ physics. First, different micro states described by, or models of, B can correspond to a single macro state/model of $E$ an example is the way in which a number of different micro states described by statistical mechanics correspond to a single macro state in thermodynamics. Second, different micro theories can correspond to the same macro theory. An example is how fluids of different micro-constitutions (i.e., cells, molecules, atoms, or particles of different types) can give rise to the same hydrodynamic behaviour at a coarser-grained description. 
	
	This conception of emergence can be used to understand hierarchical emergence in several different approaches to QG \citep{Crowther2016}. Here, I present just two of these: \textit{causal set theory} \S\ref{CST}, and \textit{loop quantum gravity} (LQG), \S\ref{LQG}. Before this exposition, however, I briefly comment on another class of approaches towards QG in which the heirarchical conception of spacetime emergence is applicable under certain circumstances: \textit{holographic scenarios for gravity}, \S\ref{holog}
	
	\subsection{Holographic scenarios for gravity}\label{holog}
	
	There has been much discussion of emergent spacetime associated with holographic dualities. Since it is not immediately obvious which (if either) type of emergence is associated with dualities, it is worth briefly commenting on this type of QG approach.\footnote{\label{fn:holog}See, e.g., \cite{tHooft, Verlinde2011, Rickles2013, Teh2013, Dieks2015, deHaro2017, Vistarini2017}.} My claim here is simply that, if there are cases of emergent spacetime associated with holographic scenarios, these would be examples of hierarchical emergence.
	
	A \textit{duality} in physics is typically defined as an equivalence between two theories which agree on the values of all physical quantities, but may otherwise have very different formulations.\footnote{This is a loose characterisation, and one may debate its adequacy in general (particularly regarding the idea of ``physical quantities''). Questions regarding the specific nature of dualities are beyond the scope of this paper, but interested readers are encouraged to dig into the philosophical literature, including \cite{Butterfield2018, Dawid2017, Read2018} and the special issue \cite{Castellani2017}.} In a \textit{holographic} duality, a gravitational theory of (d+1)-dimensions is equivalent to a d-dimensional theory of a many body system without gravity. In order to ground any claim of emergent spacetime from the lower-dimensional theory, there must be an argument for considering the gravitational theory \textit{Dependent} on (or otherwise \textit{less fundamental} than) the lower-dimensional one. Such an argument is in conflict, however, with the equivalence suggested by the duality in the case of exact duality. Thus---as has already been well-noted in the philosophical literature---exact duality is in tension with any claim of emergence. This is the case, in particular, with the AdS/CFT duality: a holographic duality between a string theory featuring gravity, describing closed strings propagating on a spacetime (anti-de Sitter space (AdS)), known as the ``bulk'', and a gauge theory without gravity (a conformal field theory (CFT)), defined on the boundary that contains the bulk spacetime. Since the AdS/CFT duality is supposed to be exact, there is no compelling argument to be made that spacetime emerges from the CFT.\footnote{Cf. the philosophy references in Fn. \ref{fn:holog}.}
	
\begin{figure}[!htb]
		\center{\includegraphics[scale=0.6]
			{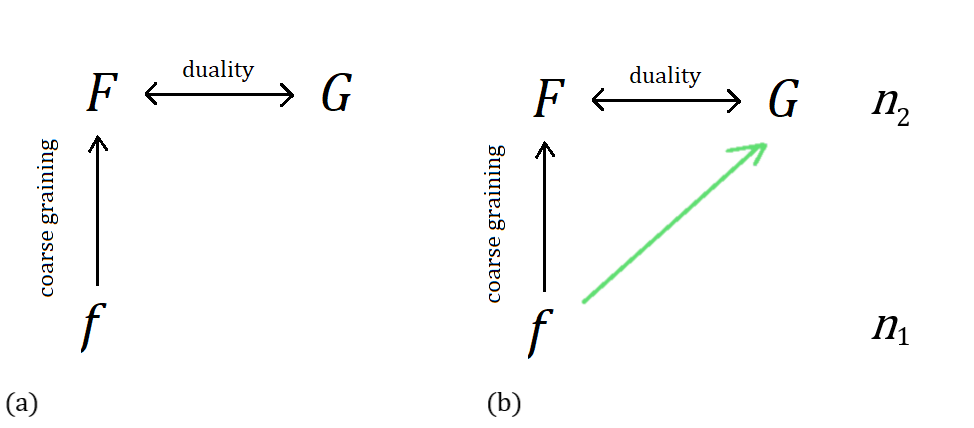}}
		\caption{(\textit{BrokenMap}): \cite{Dieks2015, deHaro2017} suggest that one way of breaking the holographic duality between theories $F$ and $G$ is if the duality only holds at the macro-level, $n_2$, but not the micro-level, $n_1$; this would be the case if $F$ has a micro-level theory, $f$, but $G$ does not have a micro-level theory. (a) schematically illustrates this scenario; (b) shows \cite{Dieks2015} suggestion that the coarse-graining relation between $f$ and $F$ can be considered together with the holographic duality between $F$ and $G$ to get an asymmetric relation (diagonal arrow) between $f$ and $G$. This arrow could potentially represent emergence, in which case it would be hierarchical emergence.}
		\label{fig:brokenmap}
	\end{figure}

	However, following \cite{deHaro2017}, there are two possible ways in which we could consider emergent spacetime associated with a holographic duality: both of which necessarily involve \textit{breaking} the duality. The first way is \textit{(BrokenMap)}: here, we suppose or discover that the duality between theory $G$ (the gravitational theory) and theory $F$ (the field theory without gravity) breaks down at some level of fine graining. This might occur, for instance, if theory $G$ is not defined at the micro-level, while theory $F$ is. In such a case, as illustrated in Fig. \ref{fig:brokenmap} the duality is only a macro-level approximation that does not hold at the micro-level. \cite{deHaro2017} puts it, ``The world seen through the lens of $G$ is a hologram of limited resolution: a fine-grained description of the world would require us to use the more fundamental $F$'' (p. 119).\footnote{Note that an additional argument for $F$'s relative fundamentality needs to be provided here.} This scenario would mean there is an asymmetry between $F$ and $G$ which provides room for a claim of emergent spacetime in $G$. Note that this scenario relies crucially on a coarse-graining relation: it is this connection between micro and macro levels that underwrites the possibility of the emergence-claim. The emergent $G$ would be a coarser-grained theory than $F$ (or more strictly, $f$) that it emerges from: thus, \textit{(BrokenMap)} is a potential model for heirarchical emergence.
	
\begin{figure}[!htb]
		\center{\includegraphics[scale=0.6]
			{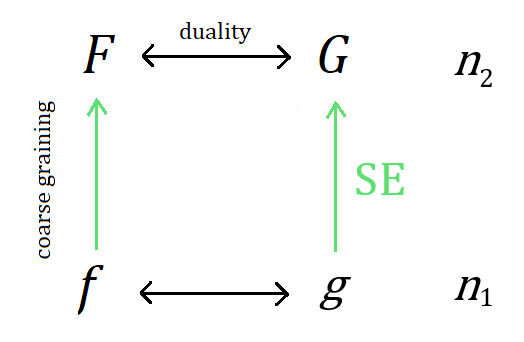}}
		\caption{(\textit{Approx}): \cite{deHaro2017} suggest that one way of breaking the holographic duality between theories $F$ and $G$ is if we apply an approximation scheme on each side of the duality. This can potentially yield a heirarchy of emergent approximate duality relations, as well as emergent structures in the higher-level ($n_2$) theories $F$ and $G$ from their basis theories $f$ and $g$ holding at the micro-level, $n_1$. This would represent hierarchical (synchronic) emergence, SE.}
		\label{fig:approx}
	\end{figure}

	The second way that \cite{deHaro2017} says we can break the exact holographic duality in order to provide the possibility of emergent spacetime is via \textit{(Approx)}. In this case, an approximation scheme (typically involving a limit) is applied on each side of the duality, so that emergent structures and quantities can arise in the coarse-grained theory, $F$ or $G$, that do not feature in the fine-grained theory, $f$ or $g$. The relation of between $g$ and $G$ that could represent emergence is independent of the duality relations between $f$ and $g$, and between $F$ and $G$, but this schema could demonstrate a hierarchy of dualities which hold approximately at each level. This is illustrated in Fig. \ref{fig:approx}. Again, the possibility of emergence crucially relies on the coarse graining procedure relating the different levels---thus, this schema provides a means of hierarchical emergence of spacetime related to holographic duality.
	
	However, while a holographic QG approach fitting the schema of either \textit{(BrokenMap)} or \textit{(Approx)}, provides the possibility for a claim of (hierarchically) emergent spacetime, it is no guarantee that an approach fitting either schema actually describes emergent spacetime. The \textit{Dependence}, \textit{Novelty}, and \textit{Autonomy} conditions would still have to be met.\footnote{While the \textit{Dependence} condition is apparently generally satisfied in the \textit{(Approx)} schema, via the coarse graining, for example, it is less-obvious that this is the case for approaches that fit \textit{(BrokenMap)}.} Evaluation of these conditions requires detailed consideration of the particular proposals that fit each of the schemas---which I do not do here. \cite{Dieks2015}, and \cite{deHaro2017} argue that the emergent gravity approach of \cite{Verlinde2011} fits the \textit{(BrokenMap)} schema for emergence, and that the AdS/CFT duality could fit the \textit{(Approx)} schema. Rather than assess these claims in any more detail, though, I turn to my other two case-studies for heirarchical emergence: causal set theory and LQG.

	\subsection{Spacetime emergence from causal set theory}\label{CST}
	
	The idea behind causal set theory is to describe spacetime as fundamentally a discrete structure---i.e., one that is composed of pointlike elements. The fundamental structures are causal sets, which provide a finer-grained description of spacetime. The elements are related only through a partial ordering (i.e., one that is transitive, antisymmetric, and reflexive) corresponding to a notion of causality. This is motivated by a theorem due to \cite{Malament1977} which states that causal structure is enough to capture spacetime geometry in GR.\footnote{Other powerful theorems to this effect are shown by \cite{Hawking1976,Levichev1987}).} A causal set is depicted as a graph of nodes and links where the nodes represent events, and the links (edges) connecting the points represent the causal relations (i.e., the partial ordering). A causal set can be constructed by working backwards: starting from a continuum spacetime, and discretising it via a `sprinkling' of points, related through the partial ordering which captures the causal (lightcone) structure of the spacetime. This `sprinkling' technique also ensures that the Lorentz invariance of the classical spacetime can be reobtained (e.g., if the graph were a regular lattice structure, then the associated spacetime would not be Lorentz invariant).
	
	Going the other way, and recovering a relativistic spacetime from a causal set, however, is more difficult, as most causal sets do not give rise to manifold-like spacetimes. A dynamics is sought that `grows' the causal sets that do give rise to relativistic spacetimes, via a `birthing' process, whereby the points in the graph successively come into existence (but note that this is not supposed to be a process \textit{in time}, and the order in which the points are born is not supposed to be physically meaningful). The dynamics of the theory is meant to ensure, too, that a given causal set corresponds to only a single spacetime. It is, however, permissible for a single spacetime to correspond to a number of distinct causal sets. \citet{WuthrichForthcoming} calls this the ``unique realisation requirement''---that the relation has to be many-to-one, not many-to-many. It is conjectured that the theory satisfies this requirement, though this has not been demonstrated. Finally, note that, currently, causal set theory is a classical approach to QG: it does not incorporate quantum mechanics.\footnote{For more on causal set theory accessible for philosophers, see \cite{Dowker2005,Henson2009,Sorkin2005,Wuthrich2012}; for a review, see \cite{Sury2019}.} The dynamics of the theory is discussed in a little more detail in \S\ref{CScosmo}, below. 
	
	For now, we just assume that causal set theory, as sketched here, is roughly correct, and treat it as a serious contender for QG. This means assuming that GR spacetime is approximately and appropriately derivable from causal set theory, even though this has not yet been demonstrated, due to the incompleteness of the approach. Thus, we assume that the \textit{Dependence} condition for hierarchical emergence is satisfied (already we can see traces of how spacetime depends on the structure of the causal set, e.g., the fact that improper `sprinkling' of points in the causal set can result in a spacetime that is not Lorentz invariant). 
	
	The \textit{Novelty} condition of emergence is clear: causal sets differ remarkably from spacetime. This is emphasised by \cite[][\S2.2]{Huggett2013}: first, there is nothing on the fundamental level corresponding to lengths and durations, second, the theory lacks the structure to identify `space' in the sense of a spacelike hypersurface, and third, there is a tension between the discreteness of the causal set and the Lorentz invariance demanded of the emergent spacetime that threatens to render the intermediate physics non-local in a way unfamiliar to GR (this last suggestion is made in \cite{Sorkin2009}).
	
\begin{figure}[!htb]
		\center{\includegraphics[scale=0.5]
			{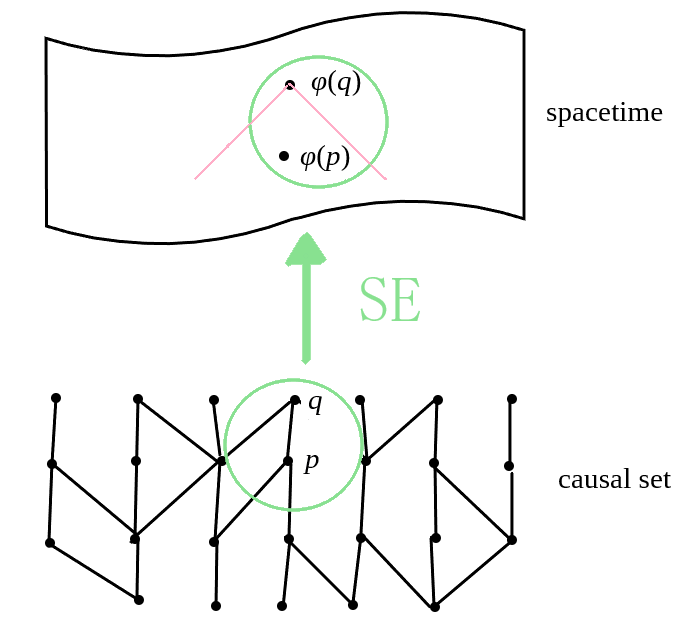}}
		\caption{Hierarchical (synchronic) emergence of spacetime from causal set theory. A relativistic spacetime corresponds to a causal set if there is a map, $\phi$, from the elements of the causal set to spacetime events which preserves the causal structure: if an elementary event $p$ causally precedes another, $q$, in the causal set, then $\phi(p)$ in the spacetime must be in the causal past lightcone (pink) of $\phi(q)$. Figure adapted from \citet{WuthrichForthcoming}.}
		\label{fig:CST}
	\end{figure}

	Finally, the \textit{Autonomy} condition of emergence is plausibly satisfied because the spacetime is robust against certain changes in the causal set, such that many different causal sets can correspond to the same spacetime. Additionally, \cite[][\S2.5]{Dowker2005} describes a potential quantum interpretation of causal set theory, using a coarse graining procedure that is importantly random, such that a spacetime could correspond dynamically to a set of `micro' states which are causal sets with no continuum approximation at all, but which have a common coarse graining to which the spacetime is a good approximation. In this case, the spacetime could be seen as autonomous from any given causal set upon which it nevertheless depends. 
	Given these suggestions, there is a plausible case to be made that relativistic spacetime can be interpreted as hierarchically emergent from the fundamental structures described by causal set theory. This is illustrated in Fig. \ref{fig:CST}.
	
	\subsection{Spacetime emergence from loop quantum gravity}\label{LQG}    
	
	Loop Quantum Gravity (LQG) is an attempt to construct a theory of QG by quantising GR. It proceeds by casting GR into Hamiltonian form, and then quantising it using the canonical quantisation procedure. This involves setting up, and solving, the theory as constraint equations, which are framed in terms of \textit{loop variables}. An intermediate step in the construction of the theory, however (moving from connection field and conjugate triad to holonomies and fluxes associated with finite paths and surfaces), can be understood as a form of discretisation, which means that the resulting theory need not be seen as `merely' a quantisation of GR, but also as bearing a close resemblence to the so-called `discrete approaches' to QG (these include, e.g., causal set theory, and approaches based on Regge discretisations).\footnote{Thanks to a referee for emphasising this point.} 
	
	This step leads to the appearance of spin network states in the theory. These are eigenstates of the so-called `area' and `volume' operators, and form the basis for a `kinematical' Hilbert space. The dynamics of the theory is less fully understood, since the \textit{Hamiltonian constraint} equation, supposed to represent the `dynamics' of the theory, resists general solution.
	The spin network states are represented as graphs called spin networks, with the nodes symbolising the quantum `volumes', and the links representing quantised `surface area' of the region bounding the volumes. The volumes are supposed to correspond to discrete `chunks' of three-dimensional space, which are adjacent to one another if there is a link connecting them. Physical space is thought to be a quantum superposition of spin network states with well-behaved geometric properties.\footnote{For more on LQG, see \cite{Rovelli2004,Rovelli2014}. Note that this latter reference is much more up-to-date than the brief sketch of the kinematic aspect of the theory that I present here; in particular, it has much more detail on the dynamics of the theory, using the covariant approach to LQG.}

	Like causal set theory, LQG is still incomplete (particularly in regards to the dynamics of the theory) and it is not yet clear how spacetime is to be recovered from the fundamental structures of the theory. For now, we will just assume that the kinematical aspect of LQG just described is roughly correct, which means assuming that \textit{space} (rather than spacetime) is fundamentally constituted by a spin network. We will also take it that LQG is a serious contender for QG, and thus assume that GR is approximately and appropriately derivable from LQG. So, we assume that the \textit{Dependence} condition for hierarchical emergence is satisfied.
	
	The \textit{Novelty} condition of emergence is plausibly satisfied because the spin network states differ from space in a number of ways. Here, I briefly mention just three of these. First, the spin networks can be seen to represent new degrees of freedom, rather than simply a quantisation of the continuum fields of GR; this is thanks to the `discretisation' step mentioned above, which leads to their interpretation as discrete (piecewise-flat or piecewise-degenerate) geometries. In other words, these degrees of freedom are one step further removed from spacetime than a quantisation \citep[in the `levels' framework of][]{Oriti2018}. Second, as has been emphasised by \cite[][\S2.3]{Huggett2013} there is a particular form of ``non-locality'', where it is possible for two `chunks of space' that are adjacent in the spin network to not map to neighbouring regions in the corresponding spacetime (though this ``non-locality'' should be heavily suppressed, otherwise the particular spin network in question would correspond to a different spacetime, one which better reflects its fundamental structure). Third, space is supposed to be a quantum superposition of spin networks, so there is no clear notion of geometry at the fundamental level. 
	
\begin{figure}[!htb]
		\center{\includegraphics[scale=0.5]
			{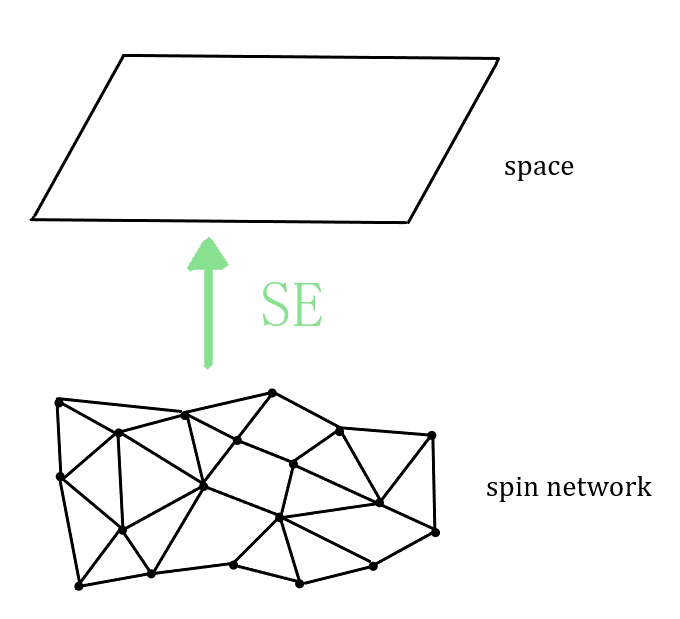}}
		\caption{Space as hierarchically (`synchronically') emergent from a spin network.}
		\label{fig:LQG}
	\end{figure}

	The \textit{Autonomy} condition of emergence is plausibly satisfied because many different spin network states can correspond to the same (semiclassical) geometry---demonstrating the robustness of the emergent spacetime. Also, given that space is meant to correspond to a superposition of spin networks, it is autonomous from any particular definite (non-superposed) spin network state. Thus, as with causal set theory, there is a plausible claim to be made that GR spacetime is emergent from the fundamental structures of LQG. This is depicted in Fig. \ref{fig:LQG} as space emergent from a definite spin network state.
	
	\section{Flat emergence of spacetime}\label{diac}
	
	The `diachronic' or `flat' conception of emergence of spacetime is supposed to describe a spatiotemporal state emergent from a `prior' non-spatiotemporal state on the same level. The flat conception of spacetime emergence is potentially applicable because one of the domains where GR is expected to be incorrect, and to require replacement by QG, is at the very beginning of the universe. Using GR and current observations of the large-scale structure of the universe, cosmologists extrapolate backwards in time in order to produce a description of the past evolution of the universe. The resulting picture (described in the direction of increasing time) is the standard, or `big bang', model of cosmology, which describes the universe expanding from a state of extremely high temperature and density approximately 13 billion years ago. Before this, however, there is the big bang singularity.
	
	One interpretation of the singular behaviour of the model is that GR is incorrect in this domain, due to its neglect of quantum effects that become important at extreme density and temperature (in which case GR likely becomes incorrect at some finite time approaching the singularity). On this view, the singularity is an unphysical artefact---a signal that GR is inapplicable here---and thus, QG should provide a correct, non-singular description of the physics in this domain.
	
	For flat emergence, $E$ and $B$ are different states of the same system, interpreted at the same level, but different times, and $E$ and $B$ are supposed to be described by different models of the same theory. A characteristic account of flat (diachronic) emergence appears in \cite{Guay2016} and \cite{Sartenaer2018}. On this account, the \textit{Dependence} condition holds that $E$ is the product of a spatiotemporally continuous process going from $B$, and/or $E$ is caused by $B$. The Novelty condition states that $E$ exhibits new entities, properties or powers that do not exist in $B$. And the \textit{Autonomy} condition states that these new entities, properties or powers are forbidden to exist in $B$ according to the laws governing $B$. 
	
	This account is not generally applicable to the case of spacetime, since it relies on spatiotemporal notions such as causation, location, and continuous processes.\footnote{Although these notions may have non-spatiotemporal analogues, e.g., causation without time \citep{Baron2014, Baron2015, Tallant2019}.} If a spatiotemporal state is to emerge from a state that is non-spatiotemporal (or, rather, less-than-fully-spatiotemporal), we cannot assume that this is a process that itself takes place in space and time (although, in fact, some approaches to QG do utilise a notion of time, this is not in all cases able to be identified with our familiar conception of time). A more general conception of flat emergence is required if we are to account for the `flat' emergence of spacetime from the `big bang' state (`big bang' is in scare quotes because this term strictly refers to the GR singularity, whereas in QG cosmology, this state may not be singular). Additionally, the \cite{Guay2016} account of emergence is a negative one, requiring that $E$ exhibit forbidden entities, properties, or powers. As explained above (\S\ref{emerg}), a negative conception of emergence is ill-suited for the case of QG, and a positive conception is to be preferred.\footnote{\cite{Shech2019} also suggests weakening the novelty condition along these lines.}
	
	The more general, positive conception of flat emergence that I propose is best-suited for understanding the flat (`diachronic') emergence of our spatiotemporal universe from a non-spatiotemporal state is one that is analogous to the hierarchical conception of emergence presented in the previous section (\S\ref{sync}).
	
	\begin{description}
		\item[EMERGENCE: FLAT CONCEPTION]
		\item[Dependence:] $E$ flatly supervenes on $B$. (Flat supervenience means that there would be no 	change in the $E$ state unless there were a change at the $B$ state, but not vice versa); 
		\item[Novelty:] $E$ differs remarkably from $B$;
		\item[Autonomy:] The physics described by $E$ is robust against changes in $B$. The `prior' state, $B$ is 	underdetermined by $E$. (This sense of underdetermination can be understood as a non-	temporal form of indeterminism, meaning that many different `initial' conditions at the $B$ 	state could give rise to the same $E$ state. If we only have knowledge of the $E$ state, this 	would not be enough information to determine the `prior' $B$ state that it `evolved from').
		
	\end{description}
	
	This account of emergence is very permissive, yet, as I demonstrate below, it is still not trivially satisfied in the case of QG cosmology.
	
	\subsection{Spacetime emergence in causal set cosmology}\label{CScosmo}
	
	The dynamics of causal set theory is the sequential growth of the causal set---i.e., the `birth' of new elements. The general class of such dynamics is known as generalised percolation, the simplest example of which is transitive percolation \citep{Rideout1999}. Here, I largely follow the exceptionally clear presentation of this by \cite{Wuthrich2017a}. Start with causal set theory's `big bang', being a single element, 1. When the next event, element 2, is birthed, it has some probability $P$ of being causally related to 1, and probability $1 - P$ of not being causally related to 1. When element 3 is birthed, it similarly has probability $P$ of being causally related to 1 (2), and probability $1 - P$ of not being causally related to 1 (2), and so on. The dynamics enforces transitive closure, so that if 1 precedes 2, and 2 precedes 3, then 1 precedes 3 (`precedes' representing the partial ordering of `causality' mentioned in \S\ref{CST} above). At any stage, the causal set that exists is called a causet. So, another way to conceive of the dynamics is that when each new causet, $\mathcal{C}'$, is born, it chooses a previously existing causet, $\mathcal{C}$, to be its ancestor with a certain probability.

\begin{figure}[!htb]
		\center{\includegraphics[scale=0.42]
			{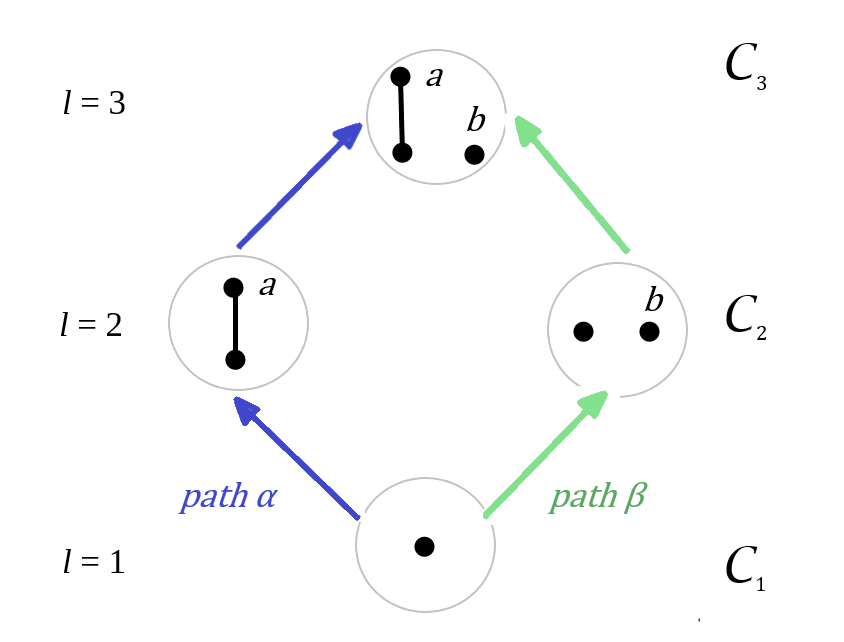}}
		\caption{\label{fig:alice}Alice and Bob's birthday parties come into being. At time $l = 2$ there is no fact of the matter whether the past followed path $\alpha$ or path $\beta$ . The causet $\mathcal{C}_2$ has two elements, but it is not determinate whether these events are 1 and $a$ (Alice's birthday), or 1 and $b$ (Bob's birthday). Figure adapted from \citet{Wuthrich2017a}.}
	\end{figure}

	As each element is birthed, it is labelled with a number, but, inspired by the general covariance of GR, this labelling is not supposed to be physically meaningful. This is called discrete general covariance, and it has the consequence that at any given stage of growth, we are in general prohibited from saying which elements exist. \citet[][919--920]{Wuthrich2017a} present a simple example illustrating this. Begin with the singleton causet, 1, at label time $l = 0$. It then births a second element, Alice's birthday, at $l = 1$, causally related to 1. Then this two-element causet births a third element, Bob's birthday, at $l = 2$, which is not causally related to 1 or 2. Call this path $\alpha$. Path $\beta$ instead birth's Bob's birthday at $l = 1$ not causally related to 1, and then Alice's birthday at  $l = 2$, causally related to 1. This is illustrated in Fig. \ref{fig:alice}.
	
	Discrete general covariance means that the product of the transition probabilities for getting from the singleton causet $\mathcal{C}_1$ to the three-element causet $\mathcal{C}_3$ is the same. This condition is used to derive the dynamics, and because the labels are pure gauge (unphysical labelling), there is no fact of the matter about which of path $\alpha$ or path $\beta$ was taken. So, in our example, there is no fact of the matter about whether Alice's birthday, $a$, or Bob's birthday, $b$, occurred in the stage immediately following the initial event.
	
	Yet, there is gauge-invariant (determinate, physical) growth: at any stage we can say how many elements are in the causet. So, at any stage of growth there is a fact of the matter as to the number of events, but there is no determinate fact regarding which events exist. Events would only have determinate existence once the entire causet has grown. The problem is that the causal sets are supposed to be future-infinite: growing `forever'. Yet, it turns out that there is a way to get determinate existence at a finite stage of growth. This is because causets based on transitive percolation will generally have many `posts'. A `post' is a single event that is comparable to every other event, i.e., an event that is either causally preceded by, or causally precedes, every other event in the causet. This may be interpreted as a cosmological model in which the universe cycles endlessly through phases of expansion, stasis, and contraction back down to a single element \citep[][024002]{Rideout1999}.
	
\begin{figure}[!htb]
		\center{\includegraphics[width=\textwidth]
			{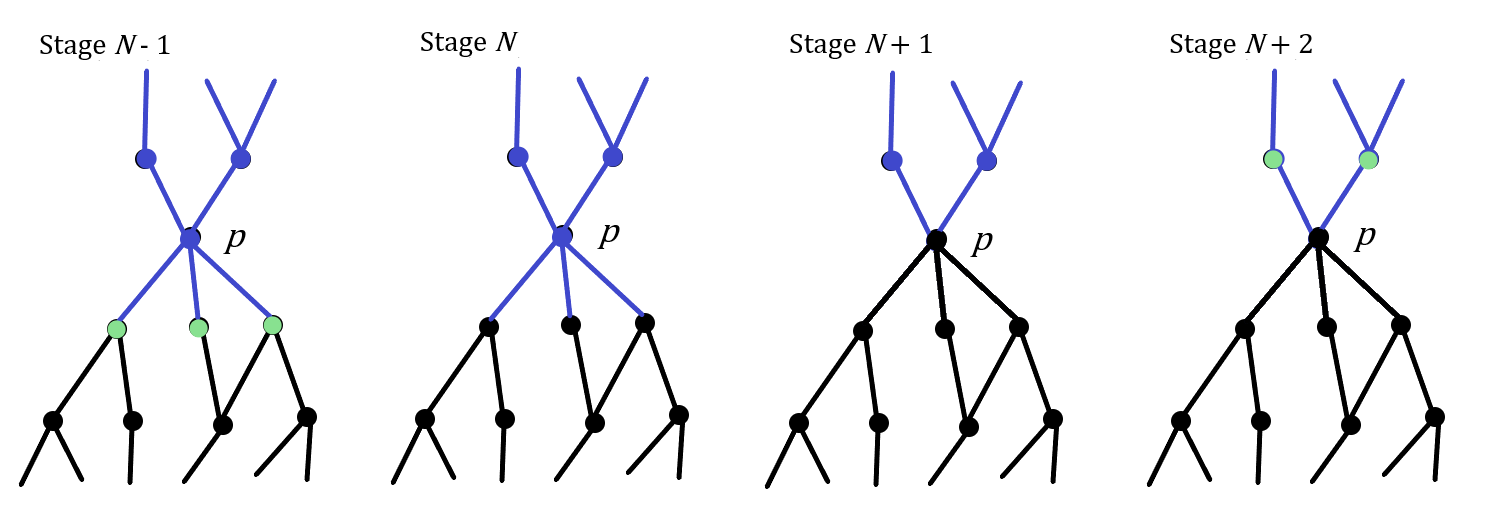}}
		\caption{\label{fig:CSCosmo}Causal set growth near a post, $p$. Events in black exist determinately, events in green exist indeterminately, and events in blue do not yet exist. Figure adapted from \citet{Wuthrich2017a}.}
	\end{figure}

	Consider an example, again from \citet[][922]{Wuthrich2017a}, shown in Fig. \ref{fig:CSCosmo}: We have a causet with a post $p$, such that N events precede $p$, and all others (infinitely many) are preceded by $p$. At stage  $l = N - 1$, (shown on the far left), the post has not yet been birthed (so it is shown in blue). At this stage, there exist $N - 1$ events, and $N - 3$ of these events determinately exist (shown in black): i.e., all the events that exist do so determinately, except the three immediately preceding $p$. Of the three events immediately preceding $p$, two must exist, but it is indeterminate which two of the three exist (i.e., these events indeterminately exist, so are shown in green). At stage $N$, immediately before $p$ is birthed, it is determinate that all $N$ ancestors of $p$ exist. There is no ontological indeterminacy at this stage, although the event $p$ itself does not yet exist. At $l = N + 1$, the post is birthed, and determinately comes into existence. At the next stage after that, $l = N + 2$, one of the two immediate successors to $p$ is birthed, but it is indeterminate which one.
	
	Now, we are interested in whether or not the expanding phase of spacetime can be said to be (flatly) emergent from the contracting phase, through the `big bang' transition state according to causal set cosmology at the micro level. Our current universe is supposed to just be one cycle of many,\footnote{In fact, causal set cosmology suggests that the universe underwent many cycles of expansion, stasis, and contraction before our `big bang' event. This is used to explain particular aspects of our current universe that are otherwise striking to cosmologists. See, \cite{Sorkin2000}.} so the event we call the `big bang' is a post with N events preceding it (several of which will also be posts), and which precedes a future-infinite number of events. Accordingly, I treat the candidate `basis' state, $B$, as the causal set depicted at Stage $N + 1$ in Fig. 4, and the candidate `emergent' state, $E$, as a generic causal set at Stage $N + n$, (where $n$ is some large but finite number), supposed to correspond to a universe in a phase of expansion. What we find is that expanding spacetime cannot be said to emerge from the contracting phase on this model at the micro level.
	
	Recall that there are three conditions that must be satisfied for emergence: \textit{Dependence}, \textit{Novelty}, and \textit{Autonomy} (\S5). In the case of this causal set cosmology model, there may be plausible claims for the \textit{Novelty} and \textit{Autonomy} of the expanding phase compared to the contracting one, but the \textit{Dependence} condition of emergence is not satisfied. This is because of the probabilistic nature of the dynamics, which ensures that the condition of flat supervenience fails: $E$ could be different without $B$ being different. This is in spite of $E$ being `causally' related to $B$, i.e., the causet $\mathcal{C}_{N+n}$ has $\mathcal{C}_{N}$ as its ancestor.
	
	Interestingly, however, there is a case to be made that the past contracting universe emerges from the expanding phase, via the big bang---or, rather `big crunch'---in its future. This is because there is a sense in which the past events depend on the big crunch event: all past events pop into determinate existence once the post is birthed. Take $B$ as the causet at Stage $N + $1 (the birthing of the post in Fig. \ref{fig:CSCosmo}), and $E$ as the causet at Stage $N$. The \textit{Dependence} condition is satisfied, since the past events depend on the determinate existence of the post for their own determinate existence. \textit{Novelty} is satisfied, because the birth of the post is special compared to births of non-post events.\footnote{We could have an even stronger basis for Novelty if we take the big bang state to be emergent compared to the state that \textit{penultimately} precedes it, i.e., the causet at Stage $N - 1$, rather than the causet at Stage $N$. In this case, the contrast would be between a state that has ontological indeterminism, versus one that has no ontological indeterminism.} The post has a novel power compared to other events: it is, in a sense, responsible for the deterministic existence of every event that precedes it.\footnote{Recall that in our positive conception of emergence, novelty is symmetric and relative: a measure of how the emergent and basis states differ from one another. Thus, \textit{Novelty} does not require that the novel power be possessed by the emergent state.} And, \textit{Autonomy} is satisfied: there is a sense in which the past events are autonomous from the post, because the causal relations run in the opposite direction (i.e., the post has the causet at Stage $N$ as its ancestor). So, while we cannot say that expanding spacetime emerges from the big bang on this model at the micro level, we \textit{can} say the converse: that the past contracting universe emerges from the big crunch in its future.
	
	\subsection{Spacetime emergence in pregeometric approaches to QG}\label{geo}
	
	\textit{Pregeometric} approaches to QG describe spacetime emergent in a phase transition. An example is \textit{quantum causal histories} (QCH), which begins with a graph resembling a causal set, but promotes the points (events) to quantum evolution operators, and attaches Hilbert spaces to the links (i.e., the causal relations) in order to make the graphs quantum-mechanical. With the Hilbert spaces on the causal relations and the events as operators, the quantum evolution strictly respects the underlying causal set. QCH is capable of modelling other pre-geometric approaches, including \textit{quantum graphity}.\footnote{For details on QCH, see, \cite{Markopoulou2009}, for quantum graphity, see: \cite{Konopka2008}.} In this approach, the dynamics is not a movement or `birthing' of points, but rather a change in the connections between the points. The connections, represented by the links of the graph, are able to be in two states `on' or `off', and, being quantum-mechanical, the generic states are superpositions of both `on' and `off'. 
	
\begin{figure}[!htb]
		\center{\includegraphics[width=\textwidth]
			{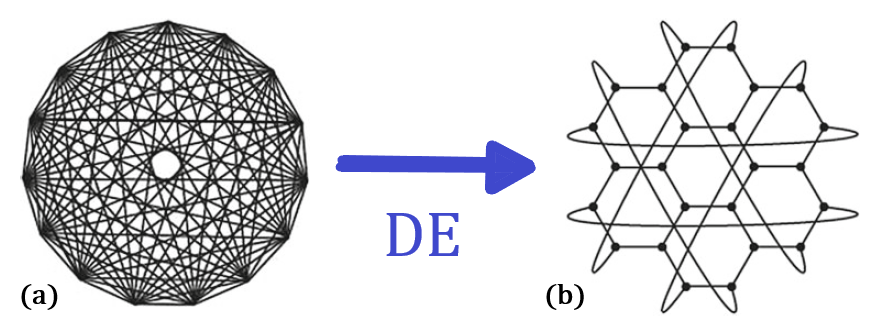}}
		\caption{Geometrogenesis as flat (diachronic) emergence. \textbf{(a)} High-energy (pre-geometric) phase of quantum graphity. \textbf{(b)} Low-energy (geometric) phase of quantum graphity. Figure adapted from \cite{Markopoulou2009}.}
		\label{fig:geo}
	\end{figure}

	The micro-description of the early (pre-geometric) universe is understood as a complete graph (as shown in Fig.\ \ref{fig:geo}a), this is a high-energy, maximally-connected state. In such a state, the dynamics is invariant under permutation of the events, and, because the entire universe is one-link adjacent to any event, there is no notion of geometry or locality. The micro-degrees of freedom are the states of the links, and these evolve in time under the Hamiltonian. As the universe cools and condenses, it undergoes a phase transition in which many of the connections switch off. The system at low-energy (i.e.\ at its ground state) is a graph with far fewer edges (Fig.\ \ref{fig:geo}b): the permutation invariance breaks, and instead translation invariance arises. At this stage locality is able to be defined and we gain a sense of relational geometry. The idea is that geometry emerges in this phase transition, known as \textit{geometrogenesis}.\footnote{Geometrogenesis is also described in \textit{group field theory}, which is an active research program that utilises, and is useful for, various other approaches to QG, including LQG and causal set theory \citep{Freidel2005, Oriti2009b, Oriti2014}.} This is illustrated in Fig. \ref{fig:geo}.
	
	Note that, in these approaches, there is a notion of time at the fundamental level, that connects the pre- and post-geometric phases. Spacetime is supposed to be associated with the geometric phase, such that the post-geometric phase is a finer-grained (lower-level) description of GR spacetime (being the higher-level phenomenon). But flat emergence concerns only a single level; here we consider the system just at the more-fundamental level of the discrete structures, rather than the `phenomenal' spacetime. So, the emergence basis $B$ is taken as the model describing the pre-geometric phase (Fig.\ \ref{fig:geo}a), and the emergent model $E$ describes the geometric phase (Fig.\ \ref{fig:geo}b).
	
	The Dependence condition can be understood as flat supervenience, since there is no change in the $E$ state unless there is a change in the $B$ state, and not vice-versa. This is ensured by the temporal aspect of these models, such that the $B$ state causally precedes the $E$ state via Hamiltonian evolution, and the two states are supposed to be of the same system, being the entire universe. The Novelty condition is satisfied given the different symmetries characterising the two states: $B$ is permutation invariant, while $E$ is not permutation invariant (but is translation invariant). Finally, the Autonomy condition is plausibly satisfied, since $E$ depends only on the broken symmetry that the system instantiates, rather than the details of $B$.\footnote{See \cite{Crowther2015, Morrison2012} for more on the relationship between symmetry-breaking and emergence.}
	
	Thus, there is a plausible sense in which spacetime could potentially be flatly emergent on these models (on which would arguably also apply to any symmetry-breaking phase transition).

	\subsection{Spacetime emergence from loop quantum cosmology}\label{LQC}
	
	Recall from \S\ref{LQG} that the dynamics of LQG is not fully understood. Loop quantum cosmology (LQC) attempts to circumvent this, by simplifying the system already at the kinematical level via the assumption that space is isotropic (the same in all directions) and homogeneous (all mass-energy is evenly distributed).\footnote{See \citet{Bojowald2011} for a technical introduction, and \citet{Huggett2018} for one aimed at philosophers.} These simplifying assumptions mean that LQC describes spatial geometry with just one degree of freedom---the scale factor operator, $\hat{p}$. Classically, the scale factor, $a$, is a variable which describes the relative `size' of space; this scale factor operator is related to the classical scale factor via $a=\sqrt{|p|}$, where the $p$s are the eigenvalues of $\hat{p}$. The physical states $\ket{\Psi}$ of the theory are those that satisfy the constraint equations, including the Hamiltonian constraint, $\hat{C}\ket{\Psi}$, which supplies the `dynamics'. These physical states are expanded in terms of some operator, such as a scalar field (though this does not itself have to be an observable). The additional degree of freedom represented by the operator can serve as `internal time' in the model, since we can study the variation of $\Psi$ with respect to this `clock'. 
	
	However, if we express the general state in the triad eigenbasis $\ket{\mu}$  of $\hat{p}$, and so use the scale factor as an internal time, the Hamiltonian constraint equation takes a particularly striking form:
	\begin{equation}\label{eq:difference}
	V_+\psi(\phi, \mu+1)+V_0\psi(\phi, \mu)+V_-\psi(\phi, \mu-1)=\hat{H}_m \psi(\phi, \mu)
	\end{equation}
	
\begin{figure}[!htb]
		\center{\includegraphics[scale=0.8]
			{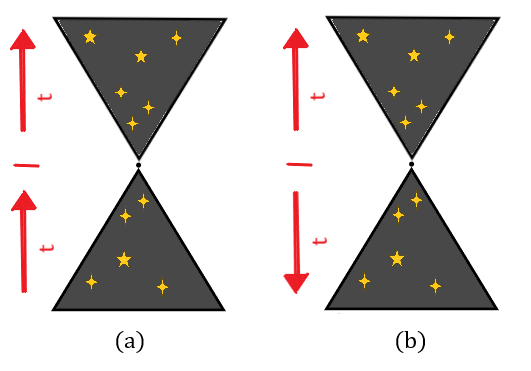}}
		\caption{Two interpretations of LQC. \textbf{(a)} The standard interpretation, as a single universe contracting (bottom triangle) then expanding (top triangle) in time (red arrow), is made difficult by the intermediate state (middle point) having no notion of time. \textbf{(b)} Huggett \& Wüthrich interpretation, as of two expanding universes that `emerge' from the single non-temporal state (point).}
		\label{fig:twinbirth}
	\end{figure}

	Where the $V$'s are coefficients to ensure the appropriate classical limit, and $\hat{H}_m$ is the matter Hamiltonian. \citet{Huggett2018} claim that (\ref{eq:difference}) can be interpreted as an evolution equation, with the scale factor operator as the `time variable' (although the difficulties with doing so are highlighted in their paper). More carefully, they would say that the universe is parameterised in a `temporal' dimension by the scale factor.\footnote{As suggested by Huggett and W\"{u}thrich in private correspondence.} The complete state of the universe can be determined using this equation, including the parts on the other side of the `big bang' at $\mu=0$. Thus, the singularity which is present in classical cosmology is not present here. The resulting picture is standardly interpreted as `big bounce', or a universe undergoing a `big crunch', contracting to a maximally hot, dense state, before re-expanding (this is depicted in Fig.\ \ref{fig:twinbirth}). 
	
	At least in one particular type of model, however, \citet{Huggett2018} argue that there is an alternative picture that is better supported by the physics. The particular class of models being referred to by \citet{Huggett2018} and \cite{Brahma2017} are only those that feature a signature change: going backward in time, from a structure that well approximates a spacetime of Lorentzian signature at late `times', to a structure with Euclidean signature in the deep quantum regime of early `times', and then back to a Lorentzian structure on the other side of the `big bang' (note that this does not correspond to the entireity of LQC, but only these particular models). In this case, there is no continuous notion of time that runs from the `pre big bang' universe through to the `post big bang' universe.\footnote{In other LQC models, however, there may be a notion of continuous evolution with respect to the scalar field as `internal time', and arguably the `big bounce' picture is better-supported.} Rather, there is an intermediate structure that divides these two phases, and this is purely spatial, with no connected notion of time at all.\footnote{Cf. \cite{Barrau}} Because of this, \cite{Huggett2018} argue that, since it more natural to interpret time as directed away from the big bang in both cases, this model could represent the ``twin birth of two universes'' from a single non-temporal state (Fig.\ \ref{fig:twinbirth}b).
	
	The model describes a situation that can be interpreted in two different ways, as shown in Fig. \ref{fig:twinbirth}. On the first interpretation (a), we could ask (i) whether the non-temporal state emerges from the contracting universe, or (ii) whether the expanding universe emerges from the non-temporal state. Thus, this first interpretation (a) offers (i) the possibility of `spacetime dissolution' (the emergence of a non-spatiotemporal state from a spatiotemporal one), as well as (ii) the possibility of spacetime emergence. But in the second option (b), we would only ask whether the two expanding universes emerge from the non-temporal state. 
	
	\cite{Brahma2017} argues that the transition from the purely spatial state to the spatiotemporal one in this model represents ``the emergence of time in loop quantum gravity'', while \cite{Huggett2018} claim that it represents the ``(a)temporal emergence of spacetime'' (though, of course, being careful to note that this is not actually supposed to be a temporal transition). Yet, neither of these papers are explicit about what they mean by `emergence'. I attempt to clarify this using the very permissive account of flat emergence developed above. 
	
	Recall that for flat emergence, we consider the system just at one level; I start by considering the `micro' level. This is made difficult in the case of LQC: the simplifying assumptions of homogeneity and isotropy that allow us to study only a single variable mean that there are no spin networks in LQC. If we want to imagine the `micro states' of LQC, they are basically just a single node (representing the one degree of freedom, the `size' of space).\footnote{But note that this is not entirely accurate, either, since it implies we can consider the dual of some cellular decomposition (in this case we just have a single cell).} Because of this, it is not sensible to discuss the micro states of LQC in this context. The implications for discussions of emergence are that, 1.\ flat emergence cannot be evaluated at the `micro level', and 2.\ the hierarchical conception of emergence is inapplicable in this context (unless, of course, one takes a different aspect of the theory to serve as the `micro level').\footnote{Thank you to a referee for suggesting this point.}
	
	I turn now to the `macro level' of spacetime (noting that the contrastive label itself is already misleading in this context). At this level, the \textit{Dependence} condition cannot be readily assessed. It is difficult to say which state `depends' on which, given that the same physics can support two interpretations---one in which spacetime `emerges', and one in which spacetime `dissolves'. Because of this, the Autonomy condition is likewise unable to be assessed. The Novelty condition is satisfied, however, given the different symmetries characterising the two states: the non-temporal state is Euclidean, of Galilean signature, while the spatiotemporal state has Lorentzian signature \citep{Barrau}.  Given that `dependence' and `autonomy' are basic conditions that characterise any account of emergence in philosophy (even beyond the General Conception that I introduced at the beginning of this paper) it seems that at this stage, we cannot determine whether or not spacetime emerges on this model at the macro level.
	
	Thus, contrary to the claims of emergence by \cite{Brahma2017} and \cite{Huggett2018}, we cannot say whether or not spacetime emerges on this model at this stage of development. 	However, by venturing into (even) more speculative territory, there is a way in which these models could possibly allow a conception of emergent spacetime at the macro level. This involves utilising \citet[][1201--1202]{Huggett2018} suggestion of how we could potentially conceive of a non-temporal region coming `before' a spatiotemporal one. Their idea is to extrapolate local, directed time beyond its proper domain of applicability. Then, the non-temporal region would be `before' the effective spatiotemporal region iff timelike curves in the effective spacetime can be extended to the non-temporal region in the past, but not in the future, effective direction. Even if the big bang region is timeless, these authors argue, it may be a past limit relative to time's arrow, and in this---very weak and novel---sense it could be said to come `before'. The non-temporal region would not have any temporal extent, so from the point of view of an effective temporal description, there would be an open timelike curve plus one extra `point', where the `point' represents to the whole Euclidean region (and so has structure, but not temporal structure).\footnote{Although this may be a misleading representation: the points of the time dimension at other times of course refer to spatial slices in the whole of spacetime, and this one `point' refers just to a space with one more dimension. Thanks to Nick Huggett for this clarification.} But, according to Huggett and W\"{u}thrich's suggestion, the initial point could be considered a time, from which later states evolve. They claim that this requires also that there be a dynamics in which the non-temporal region is an initial state (of course, if we are to make sense of such a claim, we require a sufficiently broad understanding of `dynamics').
	
	So, while usually there could be no deterministic evolution through the Euclidean region, precluding any dependence of a spatiotemporal state upon this region, if we follow Huggett and Wüthrich's suggestion (assuming that it makes sense), we suppose that the spatiotemporal region evolves from the non-temporal one, and thus depends on it, so the \textit{Dependence} condition would be satisfied. But, how sensitively would the spatiotemporal state depend on the initial state? Presumably, this initial state has to be special enough that it ensures the conditions of spatial homogeneity and isotropy hold, and this would indicate a sensitive dependence of the spatiotemporal state on the initial state, precluding any robustness or autonomy of the spatiotemporal state. On the other hand, there are models of cosmic inflation that could ensure these conditions hold, regardless of the particular details of the initial state. Thus, the addition of an inflation mechanism could ensure that the later state (of a spatially isotropic and homogeneous expanding universe) is reasonably robust against the details of the initial state (the `point'), such that the \textit{Autonomy} condition could plausibly be satisfied. 
	
	It must be emphasised that it is not known at this stage whether Huggett and W\"{u}thrich's suggestion for treating the non-temporal state as before the spatiotemporal one is supported by the models, nor whether the LQC models could accommodate inflation (although, there are claims that they can, \citep{Bojowald2011}). However, if we suppose that these suggestions are justified, then we could potentially say that there is a way to conceive of spacetime flatly emerging according to these models at the macro (effective spacetime) level. If these speculations turn out to be unsupported, however, then it is not clear that there is any sense to saying that spacetime flatly emerges in these models at the macro level.
	
	\section{Conclusion}
	
	The philosophy of QG is a speculative and precarious endeavour; it is also exciting and rewarding to explore how more familiar concepts from the philosophy of science might fare at the very frontier of physics, where spatiotemporal notions are threatened. I have proposed modifications of two more-standard conceptions of emergence in the philosophy of science, which could potentially be useful for understanding the emergence of spacetime from QG. The two conceptions of emergence I utilised are very permissive and general, yet, as I have shown, they are not trivially satisfied in the different examples of QG models. Nevertheless, I have demonstrated that there are several potential cases where, plausibly, spacetime could potentially be understood as emergent.
	
	\section*{Acknowledgements}
	Thanks to the participants at the conference on Diachronic Emergence in Cologne, and especially to Olivier Sartenaer and Andreas H\"{u}ttemann for organising it. Thanks to Christian W\"{u}thrich, Nick Huggett, Augustin Baas, Sam Baron, as well as audiences in Perth, Turin, Oxford, Lausanne, and Boston. Finally, thanks to the referees for \textit{Synthese} for their helpful feedback. Funding provided by the Swiss National Science Foundation (Grant No. 105212 165702).

	\bibliography{below}
\end{document}